\documentclass[12pt]{article}
\usepackage{enumerate}
\usepackage{amsfonts}
\usepackage{amssymb}

\begin{document}
\title{On entropic quantities related to the classical capacity of infinite dimensional
quantum channels}
\author{M.E.Shirokov \thanks{Steklov Mathematical Institute, 119991 Moscow,
Russia}}
\date{}
\maketitle

\section{Introduction}
In the study of the classical capacity of finite dimensional
quantum channels the three quantities play a basic role, namely,
the output entropy, its convex hull and their difference, called
the $\chi$-function. The Holevo capacity of a channel with
constraints defined by some subset of states is equal to the
maximal value of the $\chi$-function on this subset \cite{H-Sh-1}.
The proof of the existence of optimal ensembles is based on the
analysis of the convex hull of the output entropy
\cite{U},\cite{Sch-West-1}. It also makes possible to apply convex
analysis approach to the additivity problem \cite{A&B} and
provides an equivalent formulation of the strong additivity of the
Holevo capacity for two channels \cite{H-Sh-1} in terms of the
superadditivity of the convex hull of the output entropy for these
channels.

In this paper we consider generalizations of the $\chi$-function and
of the convex hull of the output entropy to the infinite dimensional
case developing results in \cite{Sh-2},\cite{H-Sh-2}.

It is shown that the $\chi$-function of an arbitrary channel is a
concave lower semicontinuous function on the whole state space
with natural chain properties (propositions 1-2), having
continuous restriction to any set of continuity of the output
entropy (proposition 7). This implies continuity of the
$\chi$-function for the Gaussian channels with the power
constraint (corollary 3 and notes below). For the $\chi$-function
the analog of Simon's dominated convergence theorem for quantum
entropy \cite{Simon} (corollary 1) is also obtained. These results
provide the proof of new version of theorem 2 in \cite{Sh-2},
where it is stated that the subadditivity of the $\chi$-function
for all finite dimensional channels implies the subadditivity of
the $\chi$-function for all infinite dimensional channels.

Since in the finite dimensional case the convex hull of the output
entropy is a continuous function it coincides with its convex
closure \cite{J&T} (lower envelope in terms of \cite{Alf}). In the
infinite dimensional case this coincidence does not hold and it
seems reasonable to consider the convex closure of the output
entropy instead of its convex hull. The explicit integral
representations of the convex closure of the output entropy of an
arbitrary infinite dimensional channel is obtained and its
properties are explored (propositions 3-6, corollary 2). The main
technical problem here is non-compactness of the state space which
makes impossible to apply the general theory of integral
representation on convex compact sets \cite{Alf},\cite{B&R}. The
main ingredient of this consideration is the criterion of
compactness of a subset of measures as well as other results
obtained in \cite{H-Sh-2}. It is shown that the convex closure of
the output entropy coincides with the convex hull of the output
entropy on the convex set of states with finite output entropy. Thus
the representation of the $\chi$-function as a difference between
the output entropy and its convex closure remains valid on this set.
Similarly to the case of the $\chi$-function, it is shown that the
convex closure of the output entropy has continuous restriction to
any set of continuity of the output entropy (proposition 7).

The obtained properties of the convex closure of the output
entropy make it possible to generalize to the infinite dimensional
case the convex duality approach to the additivity problem
proposed in \cite{A&B} (the theorem in section 6).

A very important particular case of the convex closure (= convex
hull) of the output entropy of a finite dimensional channel is the
notion of the entanglement of formation (EoF) \cite{B&Ko} of a state
in bipartite system. Indeed, the EoF coincides with the convex
closure of the output entropy of a partial trace channel from the
state space of bipartite system onto the state space of single
subsystem. It seems natural to define entanglement of formation of a
state in tensor product of two infinite dimensional systems in the
same way as the convex closure of the output entropy of a partial
trace channel. This definition guarantees such properties of the EoF
as convexity, lower semicontinuity on the whole state space and
continuity on the subsets with constrained mean energy. It is shown
that this definition coincides with the conventional definition of
the EoF considered in \cite{B&Ko},\cite{ESP} for all states in
bipartite system having marginal states with finite entropy.

\section{Preliminaries}

Let $\mathcal{H}$ be a separable Hilbert space,
$\mathfrak{B}(\mathcal{H})$ the algebra of all bounded operators
in $\mathcal{H}$, $\mathfrak{T}( \mathcal{H})$ the Banach space of
all trace-class operators with the trace norm $\Vert \cdot \Vert
_{1}$ and $\mathfrak{S}(\mathcal{H})$ the closed convex subset of
$\mathfrak{T}(\mathcal{H})$ consisting of all density operators
(states) in $\mathcal{H}$, which is complete separable metric
space with the metric defined by the norm. We shall use the fact
that convergence of a sequence of states to a \textit{state} in
the weak operator topology is equivalent to convergence of this
sequence  to this state in the trace norm \cite{D-A}.

A finite collection $\{\pi _{i},\rho _{i}\}$ of states $\rho _{i}$
with the corresponding probabilities $\pi _{i}$ is conventionally
called \textit{ensemble}. The state $\bar{\rho}=\sum_{i}\pi
_{i}\rho _{i}$ is called \textit{the average state} of the
ensemble.

We refer to \cite{Bil},\cite{Par} for definitions and facts
concerning probability measures on separable metric spaces. In
particular we denote $\mathrm{supp}(\pi)$ support of measure $\pi$
as defined in \cite{Par}. Following \cite{H-Sh-2} we consider an
arbitrary Borel probability measure $\pi$ on
$\mathfrak{S}(\mathcal{H})$ as \textit{generalized ensemble} and the
\textit{barycenter}
\[
\bar{\rho}(\pi )=\int\limits_{\mathfrak{S}(\mathcal{H})}\rho \pi
(d\rho ).
\]
of the measure $\pi$ as the average state of this ensemble. In
this notations the conventional ensembles correspond to measures
with finite support.

Denote by $\mathcal{P}$ the convex set of all probability measures
on $\mathfrak{S}(\mathcal{H})$ equipped with the topology of weak
convergence \cite{Bil} and by $\mathcal{P}_{\mathcal{A}}$ the
convex set of all probability measures with barycenters contained
in $\mathcal{A}\subseteq\mathfrak{S}(\mathcal{H})$. It is easy to
see (due to the result of \cite{D-A}) that $\pi \mapsto
\bar{\rho}(\pi )$ is a continuous mapping from $\mathcal{P}$ onto
$\mathfrak{S}(\mathcal{H})$.

We refer to \cite{Alf},\cite{J&T} for definitions and facts from
convex analysis. For reader's convenience all the necessary
information is presented in the Appendix.

In what follows $\log $ denotes the function on $[0,+\infty ),$
which coincides with the usual logarithm on $\left( 0,+\infty
\right) $ and vanishes at zero. If $A$ is a positive finite rank
operator in $\mathcal{H},$ then the entropy is defined as
\begin{equation}
H(A)=\mathrm{Tr}A\left( I\log \mathrm{Tr}A-\log A\right) ,
\label{ent}
\end{equation}
where $I$ is the unit operator in $\mathcal{H}$. If $A,B$ two such
operators then the relative entropy is defined as
\begin{equation}
H(A\,\Vert B)=\mathrm{Tr}(A\log A-A\log B+B-A)  \label{relent}
\end{equation}
provided $\mathrm{ran}A\subseteq\mathrm{ran}B$, and $H(A\,\Vert
B)=+\infty$ otherwise (throughout this paper $\mathrm{ran}$
denotes the closure of the range of an operator in $\mathcal{H}$).

This definitions can be extended to arbitrary positive $A$,$B\in
\mathfrak{T}(\mathcal{H})$ in the following way:
\[
H(A)=\lim_{n\rightarrow +\infty }H(P_{n}AP_{n});\;\quad H(A\,\Vert
B)=\lim_{n\rightarrow +\infty }H(P_{n}AP_{n}\Vert P_{n}BP_{n}),
\]
where $\left\{ P_{n}\right\}$ is an arbitrary sequence of finite
dimensional projectors monotonously increasing to the unit
operator $I$. In \cite{L} it is shown that the both sequences in
the above limit expressions are nondecreasing and that these
limits coincide with the values of the entropy and of the relative
providing by the conventional definitions.

We denote by $\sum_{i}\pi_{i}\rho_{i}$ a finite convex
decomposition as distinct from  countable decomposition
$\sum^{\sigma}_{i}\pi_{i}\rho_{i}$.

\section{The $\chi$-function}

Let
$\Phi:\mathfrak{S}(\mathcal{H})\mapsto\mathfrak{S}(\mathcal{H}')$ be
an arbitrary quantum channel. The output entropy
$H_{\Phi}(\rho)\equiv H(\Phi(\rho))$ of the channel $\Phi$ is
nonnegative lower semicontinuous concave function on the set
$\mathfrak{S}(\mathcal{H})$. For given
$\rho\in\mathfrak{S}(\mathcal{H})$ the quantity $\chi_{\Phi}(\rho)$
(the Holevo capacity of the $\{\rho\}$-constrained channel $\Phi$
\cite{H-Sh-1},\cite{Sh-2}) is defined as
\begin{equation}\label{chi-def-1}
\chi_{\Phi}(\rho)=\sup_{\sum_{i}\pi
_{i}\rho_{i}=\rho}\sum_{i}\pi_{i}H(\Phi (\rho _{i})\|\Phi(\rho)).
\end{equation}
It is shown in \cite{H-Sh-2} that
\begin{equation}\label{chi-def-2}
\chi_{\Phi}(\rho)=\sup_{\pi\in\mathcal{P}_{\{\rho\}}}\int\limits_{\mathfrak{S}(
\mathcal{H})}H(\Phi(\sigma)\Vert\Phi(\rho))\pi(d\sigma),
\end{equation}
where $\mathcal{P}_{\{\rho\}}$ is the set of all probability
measures on $\mathfrak{S}(\mathcal{H})$ with the barycenter
$\rho$, and that under the condition $H_{\Phi}(\rho)<+\infty$ the
supremum in (\ref{chi-def-2}) is achieved on some measure
supported by pure states.

\textbf{Definition 1.} \textit{A measure $\pi_{0}$ with the
barycenter $\rho_{0}$ supported by the set of pure states and such
that
$$
\chi_{\Phi}(\rho_{0})=\int\limits_{\mathfrak{S}(\mathcal{H})}H(\Phi(\sigma
)\Vert\Phi(\rho_{0}))\pi_{0}(d\sigma)
$$
is called a $\chi_{\Phi}$-optimal measure for a state $\rho_{0}$.}

Note that $H_{\Phi}(\rho)=+\infty$ does not imply
$\chi_{\Phi}(\rho)=+\infty$. Indeed, it is easy to construct
FI-channel $\Phi$ \footnote{FI-channel is defined in \cite{Sh-2} as
a channel from finite dimensional system into infinite dimensional
one.} such that $H_{\Phi}(\rho)=+\infty$ for any
$\rho\in\mathfrak{S}(\mathcal{H})$. On the other hand, by the
monotonicity property of the relative entropy \cite{L-2}
$$\sum_{i}\pi_{i}H(\Phi (\rho
_{i})\|\Phi(\rho))\leq\sum_{i}\pi_{i}H(\rho
_{i}\|\rho)\leq\log\dim\mathcal{H}<+\infty
$$
for arbitrary ensemble $\{\pi_{i},\rho_{i}\}$, and hence
$\chi_{\Phi}(\rho)\leq\log\dim\mathcal{H}<+\infty$ for any
$\rho\in\mathfrak{S}(\mathcal{H})$.

For arbitrary state $\rho$ such that $H_{\Phi}(\rho)<+\infty$ the
$\chi$-function has the following representation
\begin{equation}\label{chi-exp-1}
\chi _{\Phi}(\rho)=H_{\Phi}(\rho)-\mathrm{conv}{H}_{\Phi}(\rho),
\end{equation}
where $\mathrm{conv}{H}_{\Phi}(\rho)$ is a convex hull  of the
output entropy (see the Appendix)
\begin{equation}\label{conv-def}
\mathrm{conv}{H}_{\Phi}(\rho)=\inf_{\sum_{i}\pi_{i}\rho_{i}=
\rho}\sum_{i}\pi_{i}H_{\Phi}(\rho_{i}).
\end{equation}

In the finite dimensional case the output entropy
${H}_{\Phi}(\rho)$ and its convex hull
$\mathrm{conv}{H}_{\Phi}(\rho)$ are continuous concave and convex
functions on $\mathfrak{S}(\mathcal{H})$ correspondingly and the
representation (\ref{chi-exp-1}) is valid for all states. It
follows that in this case the function $\chi_{\Phi}(\rho)$ is
continuous and concave on $\mathfrak{S}(\mathcal{H})$.

In the infinite dimensional case the output entropy
${H}_{\Phi}(\rho)$ is only lower semicontinuous and, hence, the
function $\chi_{\Phi}(\rho)$ is not continuous even in the case of
the noiseless channel $\Phi$, for which
$\chi_{\Phi}(\rho)={H}_{\Phi}(\rho)$. But it turns out that the
function $\chi_{\Phi}(\rho)$ for arbitrary channel $\Phi$ has
properties similar to the properties of the output entropy
${H}_{\Phi}(\rho)$.

\textbf{Proposition 1.} \textit{The function $\chi_{\Phi}(\rho)$
is nonnegative concave and lower semicontinuous function on
$\mathfrak{S}(\mathcal{H})$.}

The proof of this proposition is based on the following lemma.

\textbf{Lemma 1.} \textit{Let $\{\pi_{i},\rho_{i}\}$ be an
arbitrary ensemble of $m$ states with the average state $\rho$ and
let $\{\rho_{n}\}$ be an arbitrary sequence of states converging
to the state $\rho$. There exists the sequence
$\{\pi^{n}_{i},\rho^{n}_{i}\}$ of ensemble of $m$ states such that
\[
\lim_{n\rightarrow+\infty}\pi^{n}_{i}=\pi_{i},\quad
\lim_{n\rightarrow+\infty}\rho^{n}_{i}=\rho_{i},\quad and \quad
\rho_{n}=\sum_{i=1}^{m}\pi^{n}_{i}\rho^{n}_{i}.
\]}

\textbf{Proof.} Without loss of generality we may assume that $\pi
_{i}>0$ for all $i$. Let $\mathcal{D}\subseteq \mathcal{H}$ be the
support of $\rho =\sum_{i=1}^{m}\pi _{i}\rho _{i}$ and $P$ be the
projector onto $\mathcal{D}$. Since $\rho _{i}\leq \pi
_{i}^{-1}\rho$ we have
\[
0\leq A_{i}\equiv \rho ^{-1/2}\rho _{i}\rho ^{-1/2}\leq \pi
_{i}^{-1}I,
\]
where we denote by $\rho ^{-1/2}$ the generalized (sometimes
called Moore-Penrose) inverse of the operator $\rho ^{1/2}$ (equal
0 on the orthogonal complement to $\mathcal{D}$).

Consider the sequence $B_{i}^{n}=\rho _{n}^{1/2}A_{i}\rho
_{n}^{1/2}+\rho_{n}^{1/2}(I_{\mathcal{H}}-P)\rho_{n}^{1/2}$ of
operators in $\mathfrak{B}(\mathcal{H})$. Since
$\lim_{n\rightarrow +\infty }\rho_{n}=\rho=P\rho$ in the trace
norm, we have
$$
\lim_{n\rightarrow +\infty }B_{i}^{n}=\rho ^{1/2}A_{i}\rho
^{1/2}=\rho _{i}
$$
in the weak operator topology. The last equality implies
$A_{i}\neq 0.$ Note that
$\mathrm{Tr}B_{i}^{n}=\mathrm{Tr}A_{i}\rho
_{n}+\mathrm{Tr}(I_{\mathcal{H}}-P)\rho_{n}<+\infty $ and hence
$$
\lim_{n\rightarrow
+\infty}\mathrm{Tr}B_{i}^{n}=\mathrm{Tr}A_{i}\rho
=\mathrm{Tr}\rho_{i}=1.
$$

Denote by $\rho _{i}^{n}=(\mathrm{Tr}B_{i}^{n})^{-1}B_{i}^{n}$  a
state and by $\pi _{i}^{n}=\pi _{i}\mathrm{Tr}B_{i}^{n}$ a
positive number
for each $i,$ then $\lim_{n\rightarrow +\infty }\pi _{i}^{n}=\pi _{i}$ and $%
\lim_{n\rightarrow +\infty }\rho _{i}^{n}=\rho _{i}$ in the weak
operator topology and hence, by the result in \cite{D-A}, in the
trace norm. Moreover,
$$
\sum_{i=1}^{m}\pi _{i}^{n}\rho _{i}^{n}=\sum_{i=1}^{m}\pi
_{i}B_{i}^{n}=\rho _{n}^{1/2}\rho ^{-1/2}\sum_{i=1}^{m}\pi
_{i}\rho _{i}\rho ^{-1/2}\rho
_{n}^{1/2}+\rho_{n}^{1/2}(I_{\mathcal{H}}-P)\rho_{n}^{1/2}=\rho
_{n}.\quad \square
$$

\textbf{Proof of the proposition.} Nonnegativity of the
$\chi$-function is obvious. Show first the concavity property of the
$\chi$-function. Note that for convex set of states with finite
output entropy this concavity easily follows from (\ref{chi-exp-1}).
But to prove concavity on the whole state space we will use a
different approach.

Let $\rho$ and $\sigma$ be arbitrary states. By definition for
arbitrary $\varepsilon>0$ there exist ensembles
$\{\pi_{i},\rho_{i}\}_{i=1}^{n}$ and $\{\mu _{j},\sigma
_{j}\}_{j=1}^{m}$ with the average states $\rho$ and $\sigma$
correspondingly such that $\sum_{i}\pi_{i}H(\Phi (\rho
_{i})\|\Phi(\rho))>\chi_{\Phi}(\rho)-\varepsilon$ and
$\sum_{j}\mu_{j}H(\Phi
(\sigma_{j})\|\Phi(\sigma))>\chi_{\Phi}(\sigma)-\varepsilon$.
Taking the mixture
\[
\{(1-\eta )\pi_{1}\rho_{1},...,(1-\eta
)\pi_{n}\rho_{n},\eta\mu_{1}\sigma_{1},...,\eta\mu_{m}\sigma_{m}\},\quad
\eta\in [0,1]
\]
of the above two ensembles we obtain the ensemble with the average
state $(1-\eta)\rho+\eta\sigma$. By using Donald's identity
\cite{D} we have
$$
\begin{array}{c}
\chi_{\Phi}((1-\eta)\rho+\eta\sigma)\geq(1-\eta)\sum_{i}\pi_{i}H(\Phi
(\rho
_{i})\|\Phi((1-\eta)\rho+\eta\sigma))\\\\+\eta\sum_{j}\mu_{j}H(\Phi
(\sigma_{j})\|\Phi((1-\eta)\rho+\eta\sigma))=
(1-\eta)\sum_{i}\pi_{i}H(\Phi(\rho
_{i})\|\Phi(\rho))\\\\+(1-\eta)H(\Phi(\rho)\|\Phi((1-\eta)\rho+\eta\sigma))
+\eta\sum_{j}\mu_{j}H(\Phi(\sigma_{j})\|\Phi(\sigma))\\\\ +\eta
H(\Phi(\sigma)\|\Phi((1-\eta)\rho+\eta\sigma)\geq
(1-\eta)\sum_{i}\pi_{i}H(\Phi(\rho_{i})\|\Phi(\rho))\\\\+\eta\sum_{j}\mu_{j}H(\Phi
(\sigma_{j})\|\Phi(\sigma))\geq
(1-\eta)\chi_{\Phi}(\rho)+\eta\chi_{\Phi}(\sigma)-2\varepsilon,
\end{array}
$$
where nonnegativity of the relative entropy was used. Since
$\varepsilon$ can be arbitrary small the concavity property of the
$\chi$-function is established.

To prove lower semicontinuity of the $\chi$-function we have to show
\begin{equation}  \label{chi-l-s-c}
\liminf_{n\rightarrow+\infty}\chi_{\Phi}(\rho_{n})\geq\chi _{\Phi
}(\rho_{0}).
\end{equation}
for arbitrary state $\rho_{0}$ and arbitrary sequence $\rho_{n}$
converging to this state $\rho_{0}$.

For arbitrary $\varepsilon>0$ let $\{\pi_{i},\rho_{i}\}$ be an
ensemble with the average $\rho_{0}$ such that
\[
\sum_{i}\pi_{i}H(\Phi(\rho_{i})\|\Phi(\rho_{0}))\geq
\chi_{\Phi}(\rho_{0})-\varepsilon.
\]
By lemma 1 there exists the sequence of ensembles $\{\pi_{i}^{n},
\rho_{i}^{n}\}$ of fixed size such that
\[
\lim_{n\rightarrow+\infty}\pi^{n}_{i}=\pi_{i},\quad
\lim_{n\rightarrow+\infty}\rho^{n}_{i}=\rho_{i},\quad\mathrm{and}\quad
\rho_{n}=\sum_{i=1}^{m}\pi^{n}_{i}\rho^{n}_{i}.
\]

By definition we have
\[
\begin{array}{c}
 \liminf_{n\rightarrow+\infty}\chi_{\Phi}(\rho_{n})\geq
\liminf_{n\rightarrow+\infty}\sum_{i}\pi_{i}^{n}H(\Phi(\rho_{i}^{n})\|\Phi(\rho_{n}))
\\\\\geq \sum_{i}\pi_{i}H(\Phi(\rho_{i})\|\Phi(\rho_{0}))\geq
\chi_{\Phi}(\rho_{0})-\varepsilon,
\end{array}
\]
where lower semicontinuity of the relative entropy \cite{W} was
used. This implies (\ref{chi-l-s-c}) (due to the freedom of the
choice of $\varepsilon$).$\square$

The similarity of the properties of the functions
$\chi_{\Phi}(\rho)$ and $H_{\Phi}(\rho)$ is stressed by the
following analog of Simon's dominated convergence theorem for
quantum entropy \cite{Simon}.

\textbf{Corollary 1.} \textit{Let $\rho_{n}$ be a sequence of
states in $\mathfrak{S}(\mathcal{H})$, converging to the state
$\rho$ and such that $\lambda_{n}\rho_{n}\leq\rho$ for some
sequence $\lambda_{n}$ of positive numbers, converging to $1$.
Then
$$
\lim_{n\rightarrow+\infty}\chi_{\Phi}(\rho_{n})=\chi_{\Phi}(\rho).
$$}
\textbf{Proof.} The condition $\lambda_{n}\rho_{n}\leq\rho$
implies decomposition
$\rho=\lambda_{n}\rho_{n}+(1-\lambda_{n})\rho'_{n}$, where
$\rho'_{n}=(1-\lambda_{n})^{-1}(\rho-\lambda_{n}\rho_{n})$. By
concavity of the $\chi$-function we have
$$
\chi_{\Phi}(\rho)\geq\lambda_{n}\chi_{\Phi}(\rho_{n})+
(1-\lambda_{n})\chi_{\Phi}(\rho'_{n})\geq
\lambda_{n}\chi_{\Phi}(\rho_{n}),
$$
which implies
$\limsup_{n\rightarrow+\infty}\chi_{\Phi}(\rho_{n})\leq\chi_{\Phi}(\rho)$.
This and lower semicontinuity of the $\chi$-function completes the
proof.$\square$

\textbf{Remark.} Corollary 1 provides the possibility to
approximate the value $\chi_{\Phi}(\rho)$ for \textit{arbitrary}
state $\rho$ by the sequence $\chi_{\Phi}(\rho_{n})$, where
$\rho_{n}$ is the sequence of finite rank approximations of the
state $\rho$ defined by
$\rho_{n}=\mathrm{Tr}(P_{n}\rho)^{-1}P_{n}\rho$, where $P_{n}$ is
the spectral projector of the state $\rho$, corresponding to $n$
maximal eigenvalues. This possibility is used in the proof of new
version of theorem 2 in \cite{Sh-2}.$\square$

By exploring the properties of the convex closure of the output
entropy in section 4 we will establish in section 5 the continuity
of the restriction of the $\chi$-function to any set of continuity
of the output entropy.

We shall also use the following chain properties of the
$\chi$-function.

\textbf{Proposition 2.} \textit{Let
$\Phi:\mathfrak{S}(\mathcal{H})\mapsto\mathfrak{S}(\mathcal{H}')$
and
$\Psi:\mathfrak{S}(\mathcal{H}')\mapsto\mathfrak{S}(\mathcal{H}'')$
be two channels. Then}
$$
\chi_{\Psi\circ\Phi}(\rho)\leq\chi_{\Phi}(\rho)\quad and \quad
\chi_{\Psi\circ\Phi}(\rho)\leq\chi_{\Psi}(\Phi(\rho)),\quad
\forall\rho\in\mathfrak{S}(\mathcal{H})
$$

\textbf{Proof.} The first inequality follows from the monotonicity
property of the relative entropy \cite{L-2} and (\ref{chi-def-1}),
while the second one is a direct corollary of the definition
(\ref{chi-def-1}) of the $\chi$-function.$\square$

\section{The convex closure of the output entropy}

In the finite dimensional case the output entropy is finite and
the $\chi$-function can be represented by (\ref{chi-exp-1}) as a
difference between the output entropy $H_{\Phi}(\rho)$ and its
convex hull $\mathrm{conv}H_{\Phi}(\rho)$. In this case the
function $\mathrm{conv}H_{\Phi}(\rho)$ is continuous and hence it
is closed in terms of convex analysis (see the Appendix). This
implies that the function $\mathrm{conv}H_{\Phi}(\rho)$ coincides
with the convex closure $\overline{\mathrm{conv}}H_{\Phi}(\rho)$
of the output entropy $H_{\Phi}(\rho)$.

In the infinite dimensional case the function
$\mathrm{conv}H_{\Phi}(\rho)$ is not closed even in the case of
noiseless channel $\Phi$. Indeed,
$\mathrm{conv}H_{\Phi}(\rho)=+\infty$ for any state $\rho$ with
$H_{\Phi}(\rho)=+\infty$ (see the proof of lemma 2 below), but
such a state $\rho$ can be represented as a limit of a sequence
$\rho_{n}$ of finite rank states, for which
$\mathrm{conv}H_{\Phi}(\rho_{n})=0$. It follows that
$\mathrm{conv}H_{\Phi}(\rho)$ is not lower semicontinuous.

It seems natural to suppose that in the finite dimensional case the
role of the function $\mathrm{conv}H_{\Phi}(\rho)$ is played by the
function $\overline{\mathrm{conv}}H_{\Phi}(\rho)$. The aim of this
section is to confirm this conjecture by exploring properties of the
function $\overline{\mathrm{conv}}H_{\Phi}(\rho)$ and its relation
to the $\chi$-function. First of all we will obtain an explicit
representation for $\overline{\mathrm{conv}}H_{\Phi}(\rho)$.

Consider the function
$$
\hat{H}_{\Phi}(\rho)=
\inf_{\pi\in\mathcal{P}_{\{\rho\}}}\int\limits_{\mathfrak{S}(\mathcal{H})}H_{\Phi}(\rho)
\pi(d\rho)\leq+\infty
$$
where $\mathcal{P}_{\{\rho\}}$ is the set of all probability
measures with the barycenter $\rho$. It is easy to see that
$\hat{H}_{\Phi}(\rho)\leq H_{\Phi}(\rho)$ for all states $\rho$ in
$\mathfrak{S}(\mathcal{H})$. By considering properties of the
function $\hat{H}_{\Phi}$ we will establish  that
$\hat{H}_{\Phi}=\overline{\mathrm{conv}}H_{\Phi}$ (proposition 5
below).

It was mentioned in the previous section that in the definition of
the $\chi$-function the supremum over all measures coincides with
the supremum over all measures with finite support (conventional
ensembles). In contrast to this we have the following

\textbf{Lemma 2.} \textit{The equality
$\hat{H}_{\Phi}(\rho)=\inf_{\sum_{i}\pi_{i}\rho_{i}=
\rho}\sum_{i}\pi_{i}H_{\Phi}(\rho_{i})=\mathrm{conv}H_{\Phi}(\rho)$
takes place only if either $H_{\Phi}(\rho)<+\infty$ or
$\hat{H}_{\Phi}(\rho)=+\infty$.}

\textbf{Proof.} If $H_{\Phi}(\rho)<+\infty$ then
$\chi_{\Phi}(\rho)=H_{\Phi}(\rho)-\mathrm{conv}H_{\Phi}(\rho)$. By
proposition 1 and corollary 1 in \cite{H-Sh-2} we have
$\chi_{\Phi}(\rho)=H_{\Phi}(\rho)-\hat{H}_{\Phi}(\rho)$ and hence
$\hat{H}_{\Phi}(\rho)=\mathrm{conv}H_{\Phi}(\rho)$.

If $H_{\Phi}(\rho)=+\infty$ then
$\mathrm{conv}H_{\Phi}(\rho)=+\infty$ since by general properties
of entropy the set of states with finite entropy is convex.
\cite{W}.$\square$

Lemma 2 implies that
$\hat{H}_{\Phi}(\rho)<\mathrm{conv}H_{\Phi}(\rho)$ for any state
$\rho$ such that $H_{\Phi}(\rho)=+\infty$ while
$\hat{H}_{\Phi}(\rho)<+\infty$.  Note that the set of such states
is nonempty. For example, in the case of noiseless channel $\Phi$
it is easy to see that $\hat{H}_{\Phi}(\rho)=0$ for any $\rho$,
but the set of states $\rho$ with $H_{\Phi}(\rho)<+\infty$ is the
subset of the first category of $\mathfrak{S}(\mathcal{H})$
\cite{W}.

Our first goal is to show that the infimum in the definition of the
$\hat{H}_{\Phi}(\rho)$ can be taken over all measures supported by
the set of pure states. For this purpose it is useful to consider
the following partial order on the set $\mathcal{P}$. Denote by
$\mathcal{S}$ the set of all convex continuous bounded function on
$\mathfrak{S}(\mathcal{H})$. We say that $\mu\succ\nu$ if and only
if
$$
\int\limits_{\mathfrak{S}(\mathcal{H})}f(\rho)\mu(d\rho)
\geq\int\limits_{\mathfrak{S}(\mathcal{H})}f(\rho)\nu(d\rho) \quad
\mathrm{for\;all\;}f\mathrm{\;in\;}\mathcal{S}.
$$
The partial order of this type is widely used in
\cite{Alf},\cite{B&R}, where measures on compact convex set are
considered. The compactness makes it possible to establish
antisymmetrical property of this partial order, which is not needed
in our consideration.

\textbf{Proposition 3.} \textit{For arbitrary state $\rho_{0}$ there
exist a measure $\pi_{0}$ in $\mathcal{P}_{\{\rho_{0}\}}$ supported
by pure states such that
$$
\hat{H}_{\Phi}(\rho_{0})=
\int\limits_{\mathfrak{S}(\mathcal{H})}H_{\Phi}(\rho)\pi_{0}(d\rho).
$$}
\textit{The measure $\pi_{0}$ can be chosen to be a measure with
support consisting of $n^{2}$ atoms (ensemble of $n^{2}$ pure
states) if and only if the state $\rho_{0}$ has finite rank $n$.}

\textbf{Proof.} In the proof of the theorem in \cite{H-Sh-2} it
was shown that the functional
\begin{equation}\label{entr-fun}
\pi \mapsto \int\limits_{\mathfrak{S}(\mathcal{H})} H_{\Phi}
(\rho)\pi(d\rho)
\end{equation}
is well defined and lower semicontinuous on the set $\mathcal{P}$
equipped with the topology of weak convergence. By proposition 2 in
\cite{H-Sh-2} the set $\mathcal{P}_{\{\rho_{0}\}}$ is compact. Hence
the above functional achieves its minimum on this set at some point
$\pi_{*}$, i.e.
\begin{equation}\label{opt-rel}
\hat{H}_{\Phi}(\rho_{0})=
\int\limits_{\mathfrak{S}(\mathcal{H})}H_{\Phi}(\rho)\pi_{*}(d\rho).
\end{equation}

To show that among all such measures $\pi_{*}$ there exists a
measure $\pi_{0}$ supported by pure states we will use the following
two simple properties of the introduced partial order:
\begin{enumerate}
\item \textit{Let $\{\mu_{n}\}$ and $\{\nu_{n}\}$ be
two sequences in $\mathcal{P}$ weakly converging to measures $\mu$
and $\nu$ correspondingly and such that $\mu_{n}\succ\nu_{n}$ for
all $n$. Then $\mu\succ\nu$;}

\item \textit{If $\mu\succ\nu$ then
$$
\int\limits_{\mathfrak{S}(\mathcal{H})}g(\rho)\mu(d\rho)
\geq\int\limits_{\mathfrak{S}(\mathcal{H})}g(\rho)\nu(d\rho)
$$
for every function $g$ which can be represented as a pointwise
limit of monotonous sequence of functions in $\mathcal{S}$.}
\end{enumerate}

By lemma 1 in \cite{H-Sh-2} there exists the sequence $\pi_{n}$ of
measures in $\mathcal{P}_{\{\rho_{0}\}}$ with finite supports,
weakly converging to $\pi_{*}$. Decomposing each atom of the measure
$\pi_{n}$ into pure states we obtain (as in the proof of the theorem
in \cite{H-Sh-2}) the measure $\hat{\pi}_{n}$ with the same
barycenter supported by the set of pure states. It is easy to see by
definition that $\hat{\pi}_{n}\succ\pi_{n}$. By compactness of the
set $\mathcal{P}_{\{\rho_{0}\}}$ there exists subsequence
$\hat{\pi}_{n_{k}}$ converging to some measure $\pi_{0}$ supported
by the set of pure states due to theorem 6.1 in \cite{Par}. Since
$\hat{\pi}_{n_{k}}\succ\pi_{n_{k}}$, the above property 1 of the
partial order $\succ$ implies $\pi_{0}\succ\pi_{*}$.

By lemma 4 in \cite{L} the convex function $-H_{\Phi}(\rho)$ is a
pointwise limit of the monotonous sequence of bounded continuous
functions $-H(P_{n}\Phi(\rho)P_{n})$, where $\left\{ P_{n}\right\}
$ is an arbitrary sequence of finite dimensional projectors
strongly increasing to the unit operator $I$. By noting that
$H(A)=-\mathrm{Tr}A\log
A+\mathrm{Tr}A\log\mathrm{Tr}A=\mathrm{Tr}A H(A/\mathrm{Tr}A)$ we
see that the functions $-H(P_{n}\Phi(\rho)P_{n})$ are convex and
hence lie in $\mathcal{S}$ for all $n$. By the above property 2 of
the partial order $\succ$ (with $g(\rho)=-H_{\Phi}(\rho)$) and
(\ref{opt-rel}) we have
$$
\hat{H}_{\Phi}(\rho_{0})=\int\limits_{\mathfrak{S}(\mathcal{H})}H_{\Phi}(\rho)\pi_{*}(d\rho)
\geq\int\limits_{\mathfrak{S}(\mathcal{H})}H_{\Phi}(\rho)\pi_{0}(d\rho).
$$
The definition of the function $\hat{H}_{\Phi}$ implies equality
in the above inequality.

Let us prove the last statement of the proposition. If the state
$\rho_{0}$ has infinite rank then any measure in
$\mathcal{P}_{\rho_{0}}$ supported by the set of pure states has
infinite support.

Let the state $\rho_{0}$ have finite rank $n$,
$\mathcal{H}_{0}=\mathrm{supp}\rho_{0}$ be an $n$-dimensional
Hilbert space and $\Phi_{0}$ be a FI-subchannel of the channel
$\Phi$, corresponding to the subspace $\mathcal{H}_{0}$ (see
\cite{Sh-2}).

If $H_{\Phi_{0}}(\rho_{0})=H_{\Phi}(\rho_{0})<+\infty$ then by
lemma 6 in \cite{Sh-2} the function $H_{\Phi_{0}}(\rho)$ is
continuous on the compact set $\mathfrak{S}(\mathcal{H}_{0})$.
This makes it possible to apply lemma A-2 in \cite{U} to show
existence of ensemble of $(\dim\mathcal{H}_{0})^{2}$ states with
the average $\rho_{0}$, optimal in the sense of the definition  of
the function $\mathrm{conv}H_{\Phi_{0}}$, which obviously
coincides with the restriction of the function
$\mathrm{conv}H_{\Phi}$ to the subset
$\mathfrak{S}(\mathcal{H}_{0})$ of the set
$\mathfrak{S}(\mathcal{H})$. By lemma 2 the restriction of the
function $\mathrm{conv}H_{\Phi}$ to the subset
$\mathfrak{S}(\mathcal{H}_{0})$ coincides with the restriction of
the function $\hat{H}_{\Phi}$ to this subset.

If $H_{\Phi_{0}}(\rho_{0})=H_{\Phi}(\rho_{0})=+\infty$ then
$\hat{H}_{\Phi}(\rho_{0})=+\infty$ and hence any decomposition of
the state $\rho_{0}$ is optimal. To show this note first that
$H_{\Phi}(\rho_{0})=+\infty$ implies $H_{\Phi}(\sigma)=+\infty$
for arbitrary state $\sigma$ such that
$\mathrm{supp}\sigma=\mathrm{supp}\rho_{0}=\mathcal{H}_{0}$.
Indeed, for such a state $\sigma$ there is a positive number
$\lambda_{\sigma}$ such that $\lambda_{\sigma}\sigma\geq\rho_{0}$.
Nonnegativity of the relative entropy implies
$$
\lambda_{\sigma}\mathrm{Tr}\Phi(\sigma)(-\log\Phi(\sigma))\geq
\mathrm{Tr}\Phi(\rho_{0})(-\log\Phi(\sigma))\geq
\mathrm{Tr}\Phi(\rho_{0})(-\log\Phi(\rho_{0}))=+\infty.
$$
Suppose $\hat{H}_{\Phi}(\rho_{0})<+\infty$. Then there exist a
measure $\pi$ with the barycenter $\rho_{0}$ such that the
function $H_{\Phi}(\rho)$ is finite $\pi$-almost everywhere. Let
$\mathcal{F}$ be a subset of $\mathfrak{S}(\mathcal{H}_{0})$ such
that $H_{\Phi}(\rho)$ is finite on the set $\mathcal{F}$ and
$\pi(\mathcal{F})=1$. The equality
$\rho_{0}=\int_{\mathcal{F}}\rho\pi(d\rho)$ implies that the
linear hull of the set of subspaces
$\{\mathrm{supp}\rho\}_{\rho\in\mathcal{F}}$ coincides with
$\mathcal{H}_{0}$ and hence there exists a finite collection
$\{\rho_{i}\}_{i=1}^{n}$ of states in $\mathcal{F}$ such that
$\mathrm{supp}(n^{-1}\sum_{i=1}^{n}\rho_{n})=\mathcal{H}_{0}$.
Since the state $n^{-1}\sum_{i=1}^{n}\rho_{n}$ is a
\textit{finite} convex combination of the states $\rho_{i},\;
i=\overline{1,n}$ with $H_{\Phi}(\rho_{i})<+\infty$ for all
$i=\overline{1,n}$ we conclude that
$H_{\Phi}(n^{-1}\sum_{i=1}^{n}\rho_{n})<+\infty$ \cite{W}. But
this contradicts to the previous observation. $\square$

\textbf{Definition 2.} \textit{A measure $\pi_{0}$ with the
properties stated in  proposition 3 is called an
$\hat{H}_{\Phi}$-optimal measure for a state $\rho_{0}$.}

It is easy to see that the set of $\hat{H}_{\Phi}$-optimal
measures coincides with the set of $\chi_{\Phi}$-optimal measures
for arbitrary state $\rho$ such that $H_{\Phi}(\rho)<+\infty$.

The other important property of the function
$\hat{H}_{\Phi}(\rho)$ is stated in the following proposition.

\textbf{Proposition 4.} \textit{The function
$\hat{H}_{\Phi}(\rho)$ is convex and lower semicontinuous on
$\mathfrak{S}(\mathcal{H})$. The function $\hat{H}_{\Phi}(\rho)$
is closed in the sense of convex analysis (see the Appendix).}

\textbf{Proof.} To prove convexity of the function
$\hat{H}_{\Phi}(\rho)$ it is sufficient to note that
$$
\lambda\mathcal{P}_{\{\rho_{1}\}}+(1-\lambda)\mathcal{P}_{\{\rho_{2}\}}\subseteq
\mathcal{P}_{\{\lambda\rho_{1}+(1-\lambda)\rho_{2}\}}
$$
for arbitrary states $\rho_{1}$, $\rho_{2}$ and $\lambda\in[0,1]$.

Suppose that the function $\hat{H}_{\Phi}(\rho)$ is not lower
semicontinuous. This implies the existence of a sequence $\rho_{n}$
converging to some state $\rho_{0}$ and such that
\begin{equation}\label{l-s-b}
\lim\limits_{n\rightarrow+\infty}\hat{H}_{\Phi}(\rho_{n})<\hat{H}_{\Phi}(\rho_{0}).
\end{equation}
By proposition 3 for each $n=1,2,...$  there exists a measure
$\pi_{n}$ in $\mathcal{P}_{\{\rho_{n}\}}$ such that
$$
\hat{H}_{\Phi}(\rho_{n})=\int\limits_{\mathfrak{S}(\mathcal{H})}H_{\Phi}(\rho)\pi_{n}(d\rho).
$$
Let $\mathcal{A}=\{\rho_{n}\}_{n=0}^{+\infty}$ be compact subset of
$\mathfrak{S}(\mathcal{H})$. By proposition 2 in \cite{H-Sh-2} the
set $\mathcal{P}_{\mathcal{A}}$ is compact. Since
$\{\pi_{n}\}\subset\mathcal{P}_{\mathcal{A}}$ there exists
subsequence $\pi_{n_{k}}$ converging to some measure $\pi_{0}$.
Continuity of the mapping $\pi \mapsto \bar{\rho}(\pi )$ implies
$\pi_{0}\in\mathcal{P}_{\{\rho_{0}\}}$. By lower semicontinuity of
the functional (\ref{entr-fun}) we obtain
$$
\hat{H}_{\Phi}(\rho)\leq
\int\limits_{\mathfrak{S}(\mathcal{H})}H_{\Phi}(\rho
)\pi_{0}(d\rho)\leq\liminf\limits_{k\rightarrow+\infty}\int\limits_{\mathfrak{S}(\mathcal{H})}
H_{\Phi}(\rho)\pi_{n_{k}}(d\rho)=\lim\limits_{k\rightarrow+\infty}\hat{H}_{\Phi}(\rho_{n_{k}}),
$$
which contradicts to (\ref{l-s-b}).$\square$

\textbf{Proposition 5.} \textit{The function $\hat{H}_{\Phi}$
coincides with the convex closure
$\overline{\mathrm{conv}}H_{\Phi}$ of the output entropy
$H_{\Phi}$ and hence lemma 2 implies
$$
\{\overline{\mathrm{conv}}H_{\Phi}(\rho)=\mathrm{conv}H_{\Phi}(\rho)<+\infty\}
\Leftrightarrow\{H_{\Phi}(\rho)<+\infty\}.
$$}

\textbf{Proof.} By proposition 4
\begin{equation}\label{c-c-ineq}
\hat{H}_{\Phi}(\rho)\leq
\overline{\mathrm{conv}}H_{\Phi}(\rho)\leq
\mathrm{conv}H_{\Phi}(\rho)\leq H_{\Phi}(\rho),\quad
\forall\rho\in\mathfrak{S}(\mathcal{H}).
\end{equation}

By lemma 2 $\hat{H}_{\Phi}(\rho)$ coincides with
$\mathrm{conv}H_{\Phi}(\rho)$ for arbitrary state $\rho$ with
finite $H_{\Phi}(\rho)$. This and (\ref{c-c-ineq}) imply
$\hat{H}_{\Phi}(\rho)=\overline{\mathrm{conv}}H_{\Phi}(\rho)$ for
all such states.

Let $\rho$ be an arbitrary state with finite
$\hat{H}_{\Phi}(\rho)$. By lemma 3 below there exists a sequence
$\rho_{n}$ of states with finite $H_{\Phi}(\rho_{n})$ converging
to the state $\rho$ and such that
$\lim\limits_{n\rightarrow+\infty}\hat{H}_{\Phi}(\rho_{n})=\hat{H}_{\Phi}(\rho)$.
By the above observation
$\hat{H}_{\Phi}(\rho_{n})=\overline{\mathrm{conv}}H_{\Phi}(\rho_{n})$
for all $n$. Since the function
$\overline{\mathrm{conv}}H_{\Phi}(\rho)$ is closed and convex it
is lower semicontinuous \cite{J&T}. It follows
$$
\overline{\mathrm{conv}}H_{\Phi}(\rho)\leq\liminf\limits_{n\rightarrow+\infty}
\overline{\mathrm{conv}}H_{\Phi}(\rho_{n})=
\lim\limits_{n\rightarrow+\infty}\hat{H}_{\Phi}(\rho_{n})=\hat{H}_{\Phi}(\rho)
$$
This and (\ref{c-c-ineq}) imply
$\hat{H}_{\Phi}(\rho)=\overline{\mathrm{conv}}H_{\Phi}(\rho)$ for
arbitrary state $\rho$. $\square$

\textbf{Lemma 3.} \textit{For arbitrary state $\rho_{0}$ with
$\hat{H}_{\Phi}(\rho_{0})<\infty$ there exists a sequence
$\rho_{n}$ of finite rank states converging to the state
$\rho_{0}$ and such that
$$H_{\Phi}(\rho_{n})<+\infty\quad and \quad
\lim\limits_{n\rightarrow+\infty}\hat{H}_{\Phi}(\rho_{n})=\hat{H}_{\Phi}(\rho_{0}).
$$}
\textbf{Proof.} Let $\pi_{0}$ be an $\hat{H}$-optimal measure for
the state $\rho_{0}$ supported by the set of pure states
(proposition 3). Since any probability measure on the
\textit{complete} separable metric space
$\mathfrak{S}(\mathcal{H})$ is \textit{tight}
\cite{Bil},\cite{Par} for arbitrary $n\in\mathbb{N}$ there exists
compact subset $\mathcal{K}_{n}$ of
$\mathrm{Extr}(\mathfrak{S}(\mathcal{H}))$ such that
$\pi_{0}(\mathcal{K}_{n})>1-1/n$. Compactness of the set
$\mathcal{K}_{n}$ implies decomposition
$\mathcal{K}_{n}=\bigcup_{i-1}^{m(n)}\mathcal{A}^{n}_{i}$, where
$\{\mathcal{A}^{n}_{i}\}_{i=1}^{m(n)}$ is a finite collection of
disjoint measurable subsets with diameter less than $1/n$. Without
loss of generality we may assume that
$\pi_{0}(\mathcal{A}^{n}_{i})>0$ for all $i$ and $n$. By
construction \textit{compact} set $\bar{\mathcal{A}}^{n}_{i}$ lies
within some closed ball $\mathcal{B}^{n}_{i}$ of diameter $1/n$
for all $i$ and $n$.

By assumption
$$
\hat{H}_{\Phi}(\rho_{0})=
\int\limits_{\mathfrak{S}(\mathcal{H})}H_{\Phi}(\rho
)\pi_{0}(d\rho)<+\infty
$$
and hence the function $H_{\Phi}(\rho)$ is finite $\pi_{0}$-almost
everywhere. Since the function $H_{\Phi}(\rho)$ is lower
semicontinuous it achieves its \textit{finite} minimum on the
compact set $\bar{\mathcal{A}}^{n}_{i}$ of \textit{positive}
measure at some pure state
$\rho^{n}_{i}\in\bar{\mathcal{A}}^{n}_{i}$. Consider the state
$\rho_{n}=(\pi_{0}(\mathcal{K}_{n}))^{-1}\sum_{i=1}^{m(n)}\pi_{0}
(\mathcal{A}^{n}_{i})\rho^{n}_{i}$. We want to show that
\begin{equation}\label{norm-dev}
\|\rho_{n}-\rho_{0}\|_{1}\leq 3/n
\end{equation}

The state
$\hat{\rho}^{n}_{i}=(\pi_{0}(\mathcal{A}^{n}_{i}))^{-1}\int\limits_{\mathcal{A}^{n}_{i}}\rho\pi_{0}(d\rho)$
lies in $\mathcal{B}^{n}_{i}$ (see the proof of lemma 1 in
\cite{H-Sh-2}). It follows that
$\|\rho^{n}_{i}-\hat{\rho}^{n}_{i}\|_{1}\leq 1/n$. By noting that
$\pi_{0}(\mathcal{K}_{n})=\sum_{i=1}^{m(n)}\pi_{0}(\mathcal{A}^{n}_{i})$
we have
$$
\begin{array}{c}
\!\!\|\rho_{n}-\rho_{0}\|_{1}=
\|(\pi_{0}(\mathcal{K}_{n}))^{-1}\sum\limits_{i=1}^{m(n)}\pi_{0}(\mathcal{A}^{n}_{i})
\rho^{n}_{i}
-\sum\limits_{i=1}^{m(n)}\int\limits_{\mathcal{A}^{n}_{i}}\rho\pi_{0}(d\rho)-
\int\limits_{\mathfrak{S}(\mathcal{H})\backslash\mathcal{K}_{n}}\rho\pi_{0}(d\rho)\|_{1}\\\\
\leq \sum\limits_{i=1}^{m(n)}\pi_{0}(\mathcal{A}^{n}_{i})
\|(\pi_{0}(\mathcal{K}_{n}))^{-1}\rho^{n}_{i}-\hat{\rho}^{n}_{i}\|_{1}
+\|\int\limits_{\mathfrak{S}(\mathcal{H})\backslash\mathcal{K}_{n}}\rho\pi_{0}(d\rho)\|_{1}\\\\
\leq
(1-\pi_{0}(\mathcal{K}_{n}))+\sum\limits_{i=1}^{m(n)}\pi_{0}(\mathcal{A}^{n}_{i})
\|\rho^{n}_{i}-\hat{\rho}^{n}_{i}\|_{1}+
\pi_{0}(\mathfrak{S}(\mathcal{H})\backslash\mathcal{K}_{n})<3/n,
\end{array}
$$
which implies (\ref{norm-dev}).

By the choice of the states $\rho^{n}_{i}$ we have
$H_{\Phi}(\rho^{n}_{i})\leq H_{\Phi}(\rho)$ for all $\rho$ in
$\mathcal{A}^{n}_{i}$. It follows that
$$
\begin{array}{c}
\hat{H}_{\Phi}(\rho_{n})\leq
(\pi_{0}(\mathcal{K}_{n}))^{-1}\sum_{i=1}^{m(n)}\pi_{0}
(\mathcal{A}^{n}_{i})H_{\Phi}(\rho^{n}_{i})\\\\\leq
(\pi_{0}(\mathcal{K}_{n}))^{-1}\sum_{i=1}^{m(n)}
\int\limits_{\mathcal{A}^{n}_{i}}H_{\Phi}(\rho)\pi_{0}(d\rho)\\\\\leq
(\pi_{0}(\mathcal{K}_{n}))^{-1}\int\limits_{\mathfrak{S}(\mathcal{H})}
H_{\Phi}(\rho)\pi_{0}(d\rho)=(\pi_{0}(\mathcal{K}_{n}))^{-1}\hat{H}_{\Phi}(\rho_{0}).
\end{array}
$$
This implies
$\limsup_{n\rightarrow+\infty}\hat{H}_{\Phi}(\rho_{n})\leq\hat{H}_{\Phi}(\rho_{0})$.
But $\lim_{n\rightarrow+\infty}\rho_{n}=\rho_{0}$ due to
(\ref{norm-dev}) and by proposition 4 we have
$\liminf_{n\rightarrow+\infty}\hat{H}_{\Phi}(\rho_{n})\geq\hat{H}_{\Phi}(\rho_{0})$.
Hence there exists
$\lim_{n\rightarrow+\infty}\hat{H}_{\Phi}(\rho_{n})=\hat{H}_{\Phi}(\rho_{0})$.

By the construction the state $\rho_{n}$ (for all $n$) is a finite
convex combination of pure states $\rho_{i}^{n}$ with finite
output entropy $H_{\Phi}(\rho_{i}^{n})$. By general properties of
entropy \cite{W} it follows that $H_{\Phi}(\rho_{n})<+\infty$ for
all $n$. $\square$

Since the set $\mathfrak{B}( \mathcal{H})$ can be identified with
the dual space for $\mathfrak{T}( \mathcal{H})$, considered as a
complex Banach space the set $\mathfrak{B}_{h}( \mathcal{H})$ of
all hermitian operators can be identified with the dual space for
real Banach space $\mathfrak{T}_{h}( \mathcal{H})$ of all
hermitian trace class operators. The nonnegative lower
semicontinuous function $H_{\Phi}(\rho)$ on
$\mathfrak{S}(\mathcal{H})$ can be extended to the lower
semicontinuous function $\bar{H}_{\Phi}(\rho)$ on
$\mathfrak{T}_{h}(\mathcal{H})$ by ascribing the value $+\infty$
to arbitrary operator in
$\mathfrak{T}_{h}(\mathcal{H})\backslash\mathfrak{S}(\mathcal{H})$.
Hence the Fenchel transform  of the function $H_{\Phi}(\rho)$ (see
the Appendix) is defined on the set
$\mathfrak{B}_{h}(\mathcal{H})$ of all hermitian operators by
\begin{equation}\label{F-transform}
 H^{*}_{\Phi}(A)=\sup_{\rho\in\mathfrak{T}_{h}( \mathcal{H})}
\left(\mathrm{Tr}A\rho-\bar{H}_{\Phi}(\rho)\right)=\sup_{\rho\in\mathfrak{S}(\mathcal{H})}
\left(\mathrm{Tr}A\rho-H_{\Phi}(\rho)\right).
\end{equation}

Double Fenchel transform $H^{**}_{\Phi}(\rho)$ is defined on the
set $\mathfrak{T}_{h}(\mathcal{H})$ by
\begin{equation}\label{FF-transform}
H^{**}_{\Phi}(\rho)=\sup_{A\in\mathfrak{B}_{h}(\mathcal{H})}
\left(\mathrm{Tr}A\rho-H^{*}_{\Phi}(A)\right).
\end{equation}

Since the function $\bar{H}_{\Phi}(\rho)$ is nonnegative its
convex closure $\overline{\mathrm{conv}}\bar{H}_{\Phi}(\rho)$
coincides with its double Fenchel transform $H^{**}_{\Phi}(\rho)$
(see the Appendix). By noting that the restriction of
$\overline{\mathrm{conv}}\bar{H}_{\Phi}(\rho)$ to the set
$\mathfrak{S}(\mathcal{H})$ coincides with
$\overline{\mathrm{conv}}H_{\Phi}(\rho)$ proposition 5 implies the
following result.

\textbf{Corollary 2.}
\textit{$\hat{H}_{\Phi}(\rho)=H^{**}_{\Phi}(\rho)=\sup\limits_{A\in\mathfrak{B}_{h}(\mathcal{H})}\;
\inf\limits_{\sigma\in\mathfrak{S}(\mathcal{H})}
\left(H_{\Phi}(\sigma)+\mathrm{Tr}A(\rho-\sigma)\right)$ for
arbitrary state $\rho\in\mathfrak{S}(\mathcal{H})$.}\vspace{5pt}

Let us consider the set
$\hat{H}_{\Phi}^{-1}(0)=\{\rho\in\mathfrak{S}(\mathcal{H})\,|\,\hat{H}_{\Phi}(\rho)=0\}$.
Note that the set
$H_{\Phi}^{-1}(0)=\{\rho\in\mathfrak{S}(\mathcal{H})\,|\,H_{\Phi}(\rho)=0\}$
is closed subset of $\mathfrak{S}(\mathcal{H})$ due to lower
semicontinuity of the quantum entropy \cite{W}.

\textbf{Proposition 6.} \textit{The set $\hat{H}_{\Phi}^{-1}(0)$
coincides with the convex closure of the set
$H_{\Phi}^{-1}(0)\cap\mathrm{Extr}\mathfrak{S}(\mathcal{H})$.}

\textbf{Proof.} Let
$\rho\in\overline{\mathrm{conv}}(H_{\Phi}^{-1}(0)\cap\mathrm{Extr}\mathfrak{S}(\mathcal{H}))$.
Then there exist a sequence of states
$\rho_{n}\in\mathrm{conv}(H_{\Phi}^{-1}(0)\cap\mathrm{Extr}\mathfrak{S}(\mathcal{H}))$
converging to $\rho$. By definition $\hat{H}_{\Phi}(\rho_{n})=0$.
Lower semicontinuity and nonnegativity of the function
$\hat{H}_{\Phi}$ (proposition 4) implies $\hat{H}_{\Phi}(\rho)=0$.

Let $\rho\in\hat{H}_{\Phi}^{-1}(0)$. By proposition 3 the state
$\rho$ is a barycenter of a particular measure $\pi_{0}$ supported
by the set of pure states and such that $H_{\Phi}(\rho)=0$ for
$\pi_{0}$-almost all $\rho$. By using arguments from the proof of
theorem 6.3 in \cite{Par} it is easy to see that this measure
$\pi_{0}$ can be approximated by the sequence of measures
$\pi_{n}$ with finite support within the set of pure states and
such that $H_{\Phi}(\rho)=0$ for $\pi_{n}$-almost all $\rho$. This
implies that for each $n$ all atoms of the measure $\pi_{n}$ are
pure states in $H_{\Phi}^{-1}(0)$. By continuity of the mapping
$\pi \mapsto \bar{\rho}(\pi)$ the state $\rho=\bar{\rho}(\pi_{0})$
is a limit of the sequence $\bar{\rho}(\pi_{n})$ of states in
$\mathrm{conv}(H_{\Phi}^{-1}(0)\cap\mathrm{Extr}\mathfrak{S}(\mathcal{H}))$.
$\square$

\section{On continuity of the functions $\chi_{\Phi}$ and  $\hat{H}_{\Phi}$}

It follows from lemma 2 that
\begin{equation}\label{chi-exp-2}
\chi _{\Phi}(\rho)=H_{\Phi}(\rho)-\hat{H}_{\Phi}(\rho),
\end{equation}
for all states with finite output entropy. This expression remains
valid in the case $H_{\Phi}(\rho)=+\infty$ if
$\hat{H}_{\Phi}(\rho)<+\infty$. Indeed, by substituting
$\hat{H}_{\Phi}$-optimal measure $\pi$ for the state $\rho$ in
expression (4) in \cite{H-Sh-2} it is easy to see that
$\chi_{\Phi}(\rho)=+\infty$.

The properties of the functions $\chi_{\Phi}$ and $\hat{H}_{\Phi}$
obtained in the previous sections allow to relate the continuity of
these functions to the continuity of the output entropy $H_{\Phi}$.

\textbf{Proposition 7.} \textit{If the restriction of the output
entropy $H_{\Phi}(\rho)$ to a particular subset
$\mathcal{A}\subseteq\mathfrak{S}(\mathcal{H})$ is continuous then
the restrictions of the functions $\chi_{\Phi}(\rho)$ and
$\hat{H}_{\Phi}(\rho)$ to the subset $\mathcal{A}$ are continuous
as well.}

\textit{Let $\{\rho_{n}\}$ be a sequence of states converging to a
state $\rho_{0}$ such that \break
$\lim_{n\rightarrow+\infty}\hat{H}_{\Phi}(\rho_{n})=\hat{H}_{\Phi}(\rho_{0})$.
Let $\pi^{*}_{n}$ be an $\hat{H}_{\Phi}$-optimal measure for the
state $\rho_{n}$ for all $n=1,2...$. Then the set of partial
limits of the sequence $\{\pi^{*}_{n}\}_{n=1}^{+\infty}$ is
nonempty and consists of $\hat{H}_{\Phi}$-optimal measures for the
state $\rho_{0}$.}

\textbf{Proof.} The first assertion of the proposition follows from
lower semicontinuity of the function $\chi_{\Phi}(\rho)$
(proposition 1), lower semicontinuity of the function
$\hat{H}_{\Phi}(\rho)$ (proposition 4) and expression
(\ref{chi-exp-2}).

Let $\{\rho_{n}\}$ and $\pi^{*}_{n}$ be sequences mentioned in the
second statement of the proposition. Since the set
$\{\rho_{n}\}_{n=0}^{\infty}$ is compact the set
$\mathcal{P}_{\{\{\rho_{n}\}_{n=0}^{\infty}\}}$ is compact by
proposition 2 in \cite{H-Sh-2}. Hence the sequence
$\pi^{*}_{n}\subseteq\mathcal{P}_{\{\{\rho_{n}\}_{n=0}^{\infty}\}}$
has partial limits. Let $\pi_{0}$ be a limit of some subsequence
$\{\pi^{*}_{n_{k}}\}_{k=1}^{+\infty}$ of the sequence
$\{\pi^{*}_{n}\}_{n=1}^{+\infty}$. By lower semicontinuity of the
functional (\ref{entr-fun}) we have
$$
\hat{H}_{\Phi}(\rho_{0})=\lim_{k\rightarrow+\infty}\hat{H}_{\Phi}(\rho_{n_{k}})=
\lim_{k\rightarrow+\infty}
\int\limits_{\mathfrak{S}(\mathcal{H})}H_{\Phi}(\rho)\pi^{*}_{n_{k}}(d\rho)\geq
\int\limits_{\mathfrak{S}(\mathcal{H})}H_{\Phi}(\rho)\pi_{0}(d\rho),
$$
which implies $\hat{H}_{\Phi}$-optimality of the measure
$\pi_{0}$. $\square$

\textbf{Corollary 3.} \textit{Let $H^{\prime }$ be a positive
unbounded operator on the space $\mathcal{H}^{\prime }$ with
discrete spectrum of finite multiplicity such that }
$$
\mathit{\mathrm{Tr}\exp (-\beta H^{\prime })<+\infty \quad for \;
all \quad \beta >0}.
$$
\textit{Then the restrictions of the functions $\chi_{\Phi}(\rho)$
and $\hat{H}_{\Phi}(\rho)$ to the subset
$\mathcal{A}_{h^{\prime}}=\{\rho\in\mathfrak{S}(\mathcal{H})\,|\,\mathrm{Tr}\,\Phi
(\rho)H^{\prime }\leq h^{\prime}\}$ are continuous for all
$h^{\prime}\geq 0$.}

\textbf{Proof.} In the proof of proposition 3 in \cite{H-Sh-2} it
is established that the condition of the corollary implies
continuity of the restriction of the function $H_{\Phi}(\rho)$ to
the subset $\mathcal{A}_{h^{\prime}}$.$\square$

As it is mentioned in \cite{H-Sh-2} the condition of corollary 3
is fulfilled for Gaussian channels with the power constraint of
the form $\mathrm{Tr}\rho H\leq h$, where $H=R^{T}\epsilon R$ is
the many-mode oscillator Hamiltonian with nondegenerate energy
matrix $\epsilon$ and $R$ are the canonical variables of the
system.

The second part of proposition 7 implies that
$\hat{H}_{\Phi}$-optimal (=$\chi_{\Phi}$-optimal) measure for
arbitrary state with finite output entropy  can be obtained as a
limit point of any sequence of $\hat{H}_{\Phi}$-optimal
(=$\chi_{\Phi}$-optimal) measures for finite rank approximations
of this state.

\textbf{Corollary 4.} \textit{Let $\rho_{0}$ be a state such that
$H_{\Phi}(\rho_{0})<+\infty$,  $P_{n}$ be a spectral projector of
$\rho_{0}$, corresponding to the maximal $n$ eigenvalues, and
$\pi^{*}_{n}$ be an $\hat{H}_{\Phi}$-optimal
(=$\chi_{\Phi}$-optimal) measure with finite support (conventional
ensemble of $n^{2}$ states) for the finite rank state
$\rho_{n}=(\mathrm{Tr}P_{n}\rho_{0})^{-1}P_{n}\rho_{0}$ for all
$n\in\mathbb{N}$. Then any partial limit of the sequence
$\{\pi^{*}_{n}\}$ is $\hat{H}_{\Phi}$-optimal
(=$\chi_{\Phi}$-optimal) measure for the state $\rho_{0}$.}

\textbf{Proof.} By using the dominated convergence theorem for
quantum entropy \cite{Simon} it is easy to see that
$$
 \lim_{n\rightarrow+\infty}H_{\Phi}(\rho_{n})=H_{\Phi}(\rho_{0}).
$$

Hence we can apply proposition 7 with
$\mathcal{A}=\{\rho_{n}\}_{n=0}^{+\infty}$. $\square$

\section{The case of tensor product}

In \cite{A&B}  the convex duality approach to the additivity problem
in the finite dimensional case was proposed. The results of previous
sections provide a generalization of this approach to the infinite
dimensional case.

For a channel $\Phi:\mathfrak{S}(\mathcal{H})\mapsto
\mathfrak{S}(\mathcal{H}^{\prime })$ and an operator $A\in
\mathfrak{B}_{+}(\mathcal{H})$ we introduce the following output
purity of the channel \cite{Sh-1}
\begin{equation}\label{nu-def}
\nu_{H}\left(\Phi ,A\right)=\inf\limits_{\rho \in
\mathfrak{S}(\mathcal{H})}\left(H_{\Phi}(\rho)+\mathrm{Tr}A\rho
\right).
\end{equation}

Note that this characteristic is a generalization of the minimal
output entropy of the channel $\Phi$ defined by
\begin{equation}\label{min-ent-def}
H_{\mathrm{min}}(\Phi)=\inf_{\rho \in
\mathfrak{S}(\mathcal{H})}H(\Phi(\rho ))=\nu_{H}\left(\Phi ,
0\right).
\end{equation}
The concavity of the quantum entropy implies that infinitum in
(\ref{nu-def}) and (\ref{min-ent-def}) can be taken over all pure
states $\rho$ in $\mathfrak{S}(\mathcal{H})$.

Let $\Psi:\mathfrak{S}(\mathcal{K})\mapsto
\mathfrak{S}(\mathcal{K}^{\prime })$ be another channel. By
considering product states it is easy to obtain the subadditivity
property of the above output purity for tensor product channel
$\Phi\otimes\Psi$ with respect to the Kronecker sum:
\begin{equation}\label{nu-subadd}
\nu _{H}\left( \Phi \otimes \Psi ,A\otimes I+I\otimes B\right)
\leq\nu _{H}\left( \Phi ,A\right) +\nu _{H}\left( \Psi ,B\right).
\end{equation}

The additivity of the minimal output entropy for the channels
$\Phi$ and $\Psi$ means \cite{H-QI}
\begin{equation}
H_{\mathrm{min}}(\Phi\otimes\Psi)=H_{\mathrm{min}}(\Phi)+H_{\mathrm{min}}(\Psi),
\label{add-min-entr}
\end{equation}
which is equivalent to equality in (\ref{nu-subadd}) with
$A=\lambda I_{\mathcal{H}}$ and $B=\mu I_{\mathcal{K}}$,
$\lambda,\mu\in \mathbb{R}$.

The Holevo capacity of the $\mathcal{A}$-constrained channel
$\Phi$ is defined by \cite{Sh-2},\cite{H-Sh-2}
\begin{equation}\label{ccap}
\bar{C}(\Phi ;\mathcal{A})=\sup_{\sum_{i}\pi _{i}\rho _{i}\in
\mathcal{A}}\sum_{i}\pi_{i}H(\Phi (\rho _{i})\|\Phi(\bar{\rho})).
\end{equation}

The strong additivity of the $\chi$-capacity for channels $\Phi$
and $\Psi$ means \cite{Sh-2}
\begin{equation}
\bar{C}\left( \Phi \otimes \Psi ;\mathcal{A}\otimes
\mathcal{B}\right) =\bar{ C}(\Phi ;\mathcal{A})+\bar{C}(\Psi
;\mathcal{B}).  \label{add-chi-cap}
\end{equation}
for arbitrary sets $\mathcal{A}\subseteq\mathfrak{S}(\mathcal{H})$
and $\mathcal{B}\subseteq\mathfrak{S}(\mathcal{K})$ such that
$H_{\Phi}(\rho)<+\infty$ for all $\rho$ in $\mathcal{A}$ and
$H_{\Psi}(\sigma)<+\infty$ for all $\sigma$ in $\mathcal{B}$.

In the finite dimensional case the strong additivity the
$\chi$-capacity for channels $\Phi$ and $\Psi$ implies additivity
of the minimal output entropy for these channels \cite{Sh-1}. But
in the infinite dimensional case this implication is an open
problem due to the existence of pure "superentangled" states,
whose partial traces have infinite entropy (see remark 3 in
\cite{Sh-2}). Denote $\omega^{\mathcal{H}
}:=\mathrm{Tr}_{\mathcal{K}}\omega$ and
$\omega^{\mathcal{K}}:=\mathrm{Tr}_{\mathcal{H} }\omega$ for
arbitrary state $\omega$ in
$\mathfrak{S}(\mathcal{H}\otimes\mathcal{K})$. \vspace{15pt}

\textbf{Theorem.} \textit{Let $\Phi
:\mathfrak{S}(\mathcal{H})\mapsto \mathfrak{S}(
\mathcal{H}^{\prime })$ and $\Psi
:\mathfrak{S}(\mathcal{K}) \mapsto
\mathfrak{S}(\mathcal{K}^{\prime })$ be arbitrary channels.
Statements (i)-(ii) are equivalent and imply (iii)-(v):}

\begin{enumerate}[(i)]
\item  \textit{For all $\omega\in\mathfrak{S}(\mathcal{H}\otimes\mathcal{K})$}
$$
\hat{H}_{\Phi\otimes\Psi}(\omega )\geq \hat{H}_{\Phi}
(\omega^{\mathcal{H}} )+ \hat{H}_{\Psi} (\omega^{\mathcal{K}});
$$\vspace{-15pt}
\item  \textit{For all $A\in \mathfrak{B}_{+}(\mathcal{H})$ and $B\in \mathfrak{B}
_{+}(\mathcal{K})$}
$$
\nu _{H}\left( \Phi \otimes \Psi ,A\otimes I+I\otimes B\right)
=\nu _{H}\left( \Phi ,A\right) +\nu _{H}\left(\Psi,B\right) ;
$$\vspace{-15pt}
\item  \textit{For all $\rho\in\mathfrak{S}(\mathcal{H})$ and $\sigma\in\mathfrak{S}(\mathcal{K})$}
$$
\hat{H}_{\Phi \otimes \Psi}(\rho\otimes\sigma)=\hat{H}_{\Phi}
(\rho)+\hat{ H}_{\Psi}(\sigma);
$$

\item \textit{The strong additivity of the $\chi$-capacity (\ref{add-chi-cap}) holds for
      the channels $\Phi$ and $\Psi$;
}
\item \textit{Additivity of the minimal output entropy (\ref{add-min-entr}) holds for
arbitrary subchannels\footnote{The notion of subchannel is defined
in \cite{Sh-2}.} $\Phi_{0}$ and $\Psi_{0}$ of the channels $\Phi$
and $\Psi$ correspondingly.}

\end{enumerate}

\textbf{Proof.}  $(\textup{i})\Leftrightarrow(\textup{ii})$ By
proposition 5 the function $\hat{H}_{\Phi}$ is the convex closure of
the function $H_{\Phi}$. The Fenchel transform $H^{\ast}_{\Phi}$ of
$H_{\Phi}$ is defined on the set $\mathfrak{B}_{h}( \mathcal{H})$ of
all hermitian operators by (\ref{F-transform}). By lemma 1 in \cite
{A&B}\footnote{Formally the considered functions do not satisfy the
condition of this lemma, but it is easy to see that all arguments in
the proof remain valid in our case.} the strong superadditivity of
the function $\hat{H}_{\Phi}$ is equivalent to the subadditivity of
the Fenchel transform $H^{\ast}_{\Phi}$ with respect to the
Kronecker sum:
\[
H^{\ast}_{\Phi}(A\otimes I_{\mathcal{K}}+I_{\mathcal{H}}\otimes
B)\leq H^{\ast}_{\Phi}(A)+H^{\ast}_{\Phi}(B),\quad \forall A\in
\mathfrak{B}_{h}( \mathcal{H}),\;\forall B\in
\mathfrak{B}_{h}(\mathcal{K}).
\]
By the definition of $H^{\ast}_{\Phi}$ the last inequality is
equivalent to
\[
\begin{array}{c}
\sup\limits_{\sigma \in \mathfrak{S}(\mathcal{H}\otimes
\mathcal{K})}\left( \mathrm{Tr}A\omega
^{\mathcal{H}}+\mathrm{Tr}B\omega^{\mathcal{K}}-H(\Phi\otimes \Psi
(\omega))\right) \\
\\
\leq \sup\limits_{\rho \in \mathfrak{S}(\mathcal{H})}\left(
\mathrm{Tr}A\rho -H(\Phi (\rho ))\right)\;+\sup\limits_{\sigma\in
\mathfrak{S}(\mathcal{K}) } \left( \mathrm{Tr}B\sigma -H(\Psi
(\sigma))\right)
\end{array}
\]
for all $A\in \mathfrak{B}_{h}(\mathcal{H})$ and $B\in
\mathfrak{B}_{h}( \mathcal{K})$.

Noting invariance of the previous inequality after changing $A$
and $B$ on $A\pm\|A\|I_{\mathcal{H}}$ and
$B\pm\|B\|I_{\mathcal{K}}$ correspondingly and using
(\ref{nu-subadd}) we obtain that $(\textup{i}
)\Leftrightarrow(\textup{ii})$.

$(\textup{i})\Rightarrow(\textup{iii})$ The inequality $"\geq"$
follows from $(\textup{i})$ while the inequality $"\leq"$ can be
deduced from the definition of the function
$\hat{H}_{\Phi\otimes\Psi}$ by considering measures on
$\mathfrak{S}(\mathcal{H}\otimes\mathcal{K})$ supported by product
states;

$(\textup{i})\Rightarrow(\textup{iv})$ It follows from theorem 1
in \cite{Sh-2};

$(\textup{i})\Rightarrow(\textup{v})$ Let $\Phi_{0}$ and
$\Psi_{0}$ be subchannels of the channels $\Phi$ and $\Psi$
corresponding to the subspaces
$\mathcal{H}_{0}\subseteq\mathcal{H}$ and
$\mathcal{K}_{0}\subseteq\mathcal{K}$. It is easy to see that
property (i) for the channels $\Phi$ and $\Psi$ implies the same
property for its subchannels $\Phi_{0}$ and $\Psi_{0}$. Due to (i)
for arbitrary state $\omega$ in
$\mathfrak{S}(\mathcal{H}_{0}\otimes\mathcal{K}_{0})$ we have
$$
H(\Phi_{0}\otimes \Psi_{0}(\omega))\geq \hat{H}_{\Phi_{0}}
(\omega^{\mathcal{H}_{0}})+\hat{H}_{\Psi_{0}}
(\omega^{\mathcal{K}_{0}})\geq
H_{\mathrm{min}}(\Phi_{0})+H_{\mathrm{min}}(\Psi_{0})
$$
This implies inequality $"\geq"$ in (\ref{add-min-entr}). Since
the converse inequality is obvious, the equality in
(\ref{add-min-entr}) is proved. $\square$

\section{On definition of the EoF}

Entanglement is a specific feature of composed quantum systems. One
of the measures of entanglement of a state of a bipartite system is
the entanglement of formation (EoF) \cite{B&Ko}.In the finite
dimensional case it is defined as
$$
E_{F}(\rho)=\min_{\sum_{i}\pi_{i}\rho_{i}=
\rho}\sum_{i}\pi_{i}H_{\Phi}(\rho_{i}),
$$
where $\Phi$ is the partial trace channel from the state space of
bipartite system onto the state space of a marginal subsystem. In
term of convex analysis this definition means that the EoF coincides
with the convex hull of the output entropy of the partial trace
channel. Continuity of the EoF established in \cite{N} implies that
it coincides with the convex closure of the output entropy of the
partial trace channel in this case.

A natural generalization of EoF to the infinite dimensional case was
considered in \cite{ESP} and it was defined  by
$$
E_{F}^{1}(\rho)=\inf_{\sum_{i}^{\sigma}\pi_{i}\rho_{i}=
\rho}\sum_{i}\pi_{i}H_{\Phi}(\rho_{i}),
$$
where infimum is over all countable decomposition of a state
$\rho$ into pure states and $\Phi$ is the partial trace channel.

An alternative approach to the definition of the EoF was considered
in \cite{M} in the case of tensor product of two systems with one of
them finite dimensional. In this spirit we can define the EoF in the
general case by
$$
 E_{F}^{2}(\rho)=\hat{H}_{\Phi}(\rho)=
\inf_{\pi\in\mathcal{P}_{\{\rho\}}}\int\limits_{\mathfrak{S}(\mathcal{H})}H_{\Phi}(\rho)
\pi(d\rho),
$$
where $\Phi$ is the partial trace channel.

Propositions 4 and 5 imply that $E_{F}^{2}$ is a convex lower
semicontinuous function which coincides with the convex closure of
the output entropy of the partial trace channel. Proposition 3 shows
that the infimum in the above expression is achieved at some measure
supported by a set of pure states while proposition 6 implies the
following natural property of $E_{F}^{2}$:
$$
\{E_{F}^{2}(\rho)=0\}\Leftrightarrow\{\mathrm{state}\;\rho\;
\mathrm{is}\; \mathrm{nonentangled}\}
$$
where the set of nonentangled states is defined as the convex
closure of all product pure states. Indeed, for the partial trace
channel $\Phi$ the set
$H_{\Phi}^{-1}(0)\cap\mathrm{Extr}\mathfrak{S}(\mathcal{H})$
coincides with the set of all product pure states. Proposition 7
and proposition 3 in \cite{ESP} implies that $E_{F}^{2}$ is trace
norm continuous on the subsets of states with constrained mean
energy.

An interesting question is the relations between $E_{F}^{1}$ and
$E_{F}^{2}$. By proposition 3 we have
$$
E_{F}^{1}(\rho)\geq E_{F}^{2}(\rho)
$$
for all states $\rho$. Since an arbitrary state can be represented
as a countable convex combination of pure states lemma 2 and
concavity of the output entropy imply
\begin{equation}\label{EoF-eq}
E_{F}^{1}(\rho)= E_{F}^{2}(\rho)
\end{equation}
for all states $\rho$ having partial traces with finite entropy.
It is easy to see that $(\ref{EoF-eq})$ obviously holds for all
nonentangled and all pure states (for which $\hat{H}_{\Phi}$
coincides with $H_{\Phi}$). Note that lemma 3 implies
$$
E_{F}^{2}(\rho)=\lim_{\varepsilon\rightarrow+0}\inf_{\sum_{i}\pi_{i}\rho_{i}\in
\mathcal{U}_{\varepsilon}(\rho)}\sum_{i}\pi_{i}H_{\Phi}(\rho_{i}),
$$
where $\mathcal{U}_{\varepsilon}(\rho)$ is a
$\varepsilon$-vicinity of the state $\rho$ and the infimum is over
all (finite) ensembles of pure states. But the validity of
equality $(\ref{EoF-eq})$ for mixed states having partial traces
with infinite entropy remains an open problem.

\section{Appendix}

Here the notions from the convex analysis used in the main text
are presented, following \cite{J&T}. Let $f$ be an arbitrary real
valued function defined on closed convex subset $X$ of some
locally convex Hausdorff topological space. Consider the subset
$\mathrm{epi}(f)=\{(x,\lambda)\in
X\times\mathbb{R}\,|\,\lambda\geq f(x)\}$ of the set
$X\times\mathbb{R}$. Note that a function $f$ is uniquely
determined by the corresponding set $\mathrm{epi}(f)$. A function
$f$ is called \textit{convex} if the set $\mathrm{epi}(f)$ is a
convex subset of $X\times\mathbb{R}$ and it is called
\textit{closed} if the set $\mathrm{epi}(f)$ is a closed subset of
$X\times\mathbb{R}$. If a function $f$ does not take the value
$-\infty$ then its convexity means that
$$
f(\lambda x_{1}+(1-\lambda)x_{2})\leq \lambda
f(x_{1})+(1-\lambda)f(x_{2}),\quad \forall x_{1},x_{2}\in X,\;
\forall \lambda\in [0,1].
$$

Each closed function $f$ is lower semicontinuous in the sense that
the set defined by the inequality $f(x)\leq\lambda$ is a closed
subset of $X$ for arbitrary $\lambda\in\mathbb{R}$ and,
conversely, each lower semicontinuous function $f$ is closed. It
is possible to show that the lower semicontinuity of a function
$f$ means that
$$
\liminf_{n\rightarrow+\infty}f(x_{n})\geq f(x_{0})
$$
for arbitrary sequence $\{x_{n}\}$ converging to $x_{0}$.

Let $f$ be an arbitrary function on $X$. The convex hull
$\mathrm{conv}f$ of the function $f$ is defined by
$$
\mathrm{conv}f(x)=\inf_{(x,\lambda)\in\mathrm{conv}(\mathrm{epi}(f))}\lambda,
$$
where the symbol $\mathrm{conv}$ in the right side means the
convex hull of a set. This is equivalent to the following
representation
$$
\mathrm{conv}f(x)=\inf_{\sum_{i}\pi_{i}x_{i}=x}\sum_{i}\pi_{i}f(x_{i}),\quad
\pi_{i}>0,\quad \sum_{i}\pi_{i}=1.
$$
It follows that $\mathrm{conv}f$ is the greatest convex function
majorized by $f$. The convex closure $\overline{\mathrm{conv}}f$
of the function $f$ is defined by
$$
\mathrm{epi}(\overline{\mathrm{conv}}f)=\overline{\mathrm{conv}}(\mathrm{epi}(f)),
$$
where the symbol $\overline{\mathrm{conv}}$ in the right side
means the closure of the convex hull of a set. It follows that
$\overline{\mathrm{conv}}f$ is the greatest convex and closed
function majorized by $f$. This impies
$$
\overline{\mathrm{conv}}f(x)\leq\mathrm{conv}f(x)\leq f(x),\quad
\forall x\in X.
$$
If $f$ is a continuous function on \textit{compact} convex set $X$
then $\overline{\mathrm{conv}}f=\mathrm{conv}f$ \cite{Alf}.

For arbitrary real valued function  $f$ on locally convex real
linear topological space $X$ the Fenchel transform $f^{*}$ is a
function on the dual space $X^{*}$ defined by
$$
f^{*}(y)=\sup_{x\in X}(\langle y,x\rangle-f(x)),\quad \forall y\in
X^{*}.
$$
The double Fenchel transform $f^{**}$ is a function on the space
$X$ defined by
$$
f^{**}(x)=\sup_{y\in X^{*}}(\langle y,x\rangle-f^{*}(y)),\quad
\forall x\in X.
$$
By Fenchel's theorem $f^{**}(x)=\overline{\mathrm{conv}}f$ for
arbitrary function $f$ which does not take the value $-\infty$.
This implies that in this case $\overline{\mathrm{conv}}f$
coincides with the upper bound of the set of all affine continuous
functions majorized by $f$. \vspace{5pt}

\textbf{Acknowledgments.} The author is grateful to A. S. Holevo
for motivation of this work and permanent help. The work was
partially supported by INTAS grant 00-738.


\begin{thebibliography}{99}

\bibitem{Alf} Alfsen E., "Compact convex sets and boundary
integrals", Springer Verlag, 1971.

\bibitem{A&B} Audenaert K.M.R., Braunstein S.L., "On strong subadditivity of
the entanglement of formation", e-print quant-ph/0303045, 2003.

\bibitem{D} Donald M.J. "Further results on the relative entropy",
Math. Proc. Cam. Phil. Soc. 101, 363-373, 1987;

\bibitem{D-A} Dell'Antonio G.F., "On the limits of sequences of normal
states", Commun. Pure Appl. Math. 20, 413-430, 1967;

\bibitem{B&R} Bratteli O., Robinson D.W., "Operators algebras and quantum statistical mechanics";
Springer Verlag, New York-Heidelberg-Berlin, vol.I, 1979.

\bibitem{Bil} Billingsley P. "Convergence of probability measures",
John Willey and Sons. Inc., New York-London-Sydney-Toronto;

\bibitem{B&Ko} Bennett C.H., DiVincenzo D.P., Smolin J.A., Wootters W.K., "Mixed State Entanglement and Quantum Error
Correction", Phys. Rev. A 54, 3824-3851, 1996, quant-ph/9604024;

\bibitem{ESP} Eisert J., Simon C., Plenio M.B., "The quantification of entanglement in infinite-dimensional
quantum systems", J. Phys. A 35, 3911 (2002), quant-ph/0112064;

\bibitem{H-QI}  Holevo, A. S.: "Introduction to quantum information
theory". Moscow Independent University, 2002 (in Russian);

\bibitem{H-c-w-c} Holevo, A. S., "Classical capacities of quntum channels with
constrained inputs", Probability Theory and Applications, 48, N.2,
359-374, 2003,  e-print quant-ph/0211170;

\bibitem{H-Sh-1}  Holevo, A.S., Shirokov M.E., "On Shor's channel extension and
constrained channels", Commun. Math. Phys., 249, 417-430, 2004;

\bibitem{H-Sh-2}  Holevo, A.S., Shirokov M.E., "Continuous ensembles and the $\chi$-capacity
of infinite dimensional channels", e-print quant-ph/0408176, 2004;

\bibitem{L} Lindblad, G., "Expectation and entropy inequalities for finite
quantum systems", Comm. Math. Phys. 39, N.2, 111-119, 1974.

\bibitem{L-2} Lindblad, G., "Completely Positive Maps and Entropy Inequalities",
Comm. Math. Phys. 40, N.2, 147-151, 1975.

\bibitem{M} Majewski A.W., "On the measure of entanglement",  J.Phys.A 35. 123,
2002, e-print quant-ph/0101030;

\bibitem{N} Nielsen M.A., Continuity bounds for entangelment, Phys. Rev. A
61, N6, 064301, 2000.

\bibitem{J&T}  Joffe, A. D., Tikhomirov, B. M.: "Theory of extremum
problems", Moscow: Nauka, 1974 (in Russian)

\bibitem{Par} Parthasarathy, K., "Probability measures on metric spaces",
Academic Press, New York and London, 1967;


\bibitem{Sch-West-1} Schumacher, B., Westmoreland, M.: "Optimal signal
ensemble", Phys. Rev. A 51, 2738, 1997;

\bibitem{Sh-2} Shirokov, M.E., "The Holevo capacity of infinite dimensional
channels."   e-print quant-ph/0408009, 2004;

\bibitem{Sh-1}  Shirokov M.E., "On the additivity conjecture for channels with
arbitrary constrains", LANL e-print quant-ph/0308168, 2003;

\bibitem{Simon}  Simon, B., "Convergence theorem for entropy", appendix in
Lieb E.H., Ruskai M.B., "Proof of the strong suadditivity of
quantum mechanical entropy", J.Math.Phys. 14, 1938, 1973;

\bibitem{W} Wehrl, A., "General properties of entropy", Rev. Mod. Phys. 50,
221-250, 1978;

\bibitem{U}  Uhlmann, A., "Entropy and optimal decomposition of
states relative to a maximal commutative subalgebra", LANL e-print
quant-ph/9704017, 1997.

\end{thebibliography}
\end{document}